\documentclass[twocolumn, prb, showpacs,superscriptaddress]{revtex4}
\usepackage{amssymb}

\usepackage{graphicx}
\usepackage{dcolumn}
\usepackage{bm}
\usepackage{amsmath}

\begin{document}

\title{Properties of A Class of Topological Phase Transition}
\author{Zi Cai}
\affiliation{ Beijing National Laboratory for Condensed Matter
Physics, Institute of Physics, Chinese Academy of Sciences, Beijing
100080, P. R. China}
\author{Shu Chen}
\affiliation{ Beijing National Laboratory for Condensed Matter
Physics, Institute of Physics, Chinese Academy of Sciences, Beijing
100080, P. R. China}
\author{Supeng Kou}
\affiliation{Department of physics, Beijing Normal University,
Beijing 100875, P. R. China}
\author{Yupeng Wang}
\affiliation{ Beijing National Laboratory for Condensed Matter
Physics, Institute of Physics, Chinese Academy of Sciences, Beijing
100080, P. R. China}

\date{Received \today }

\begin{abstract}
The properties of a class of topological quantum phase transition
(TQPT) are analyzed based on a model proposed by Haldane. We study
the effect of finite temperature on this phase transition. We have
found that finite temperature would drive this TQPT to be a
crossover, while it is stable against the weak short range
interaction. When the interaction is strong enough, however, this
TQPT is unstable and other states would emerge. Then we investigate
the effect of the on-site energy in the original haldane model. The
critical difference between our TQPT and the topological phase
transition in conventional quantum Hall system is discussed.
Finally, we discuss the potential application of our analysis to a
topological phase transition proposed in a realistic system.
\end{abstract}
 \maketitle

\section{Introduction}

Landau's theories of the continuous phase transition have been successful
and provided a paradigm in the condensed matter physics. The spontaneous
symmetry breaking plays a central role in Landau's theories. In zero
temperature however, it is possible that the change of fermi-surface
topology or the quantum fluctuation induce new kind of phase transitions
\cite{Sachdev} beyond Landau's paradigm. The different phases in these phase
transitions are not classified by different symmetries, instead, they are
characterized by different topology of Fermi surface \cite{Lifshiz, Volovik}%
, different quantum numbers \cite{Read, Bernevig} or nonlocal topological
order parameters \cite{Wen, Feng, kou} etc. This kind of phase transition,
known as topological quantum phase transition(TQPT), has attracted
considerable attention in recent years because they are closely related with
the field of Topological Quantum Computation \cite{Das}. Despite of great
efforts, many properties of the Topological phase and TQPT, such as its
order and universality classes , the effect of temperature and interaction
and its stability, are still far from being completely understood. For
example, since there is no spontaneous symmetry breaking in the TQPT, the
Mermin-Wagner Theorem\cite{Mermin} could not be applied to preclude the
possibility of a finite temperature phase transition in 2D system. In this
paper, we have partly answered above questions.

\section{The Haldane model}

In this paper, we present a theoretical analysis of a class of TQPT. Our
starting point is the Haldane model\cite{Haldane}, which was proposed to
study the realization of ''parity anomaly'' in graphene-like condensed
matter system and is the first work of anomalous Hall effect. The phase
transition in this model, essentially, is the transition between two phases
with different parity, or in another word, different Chern numbers, which is
induced by adjusting parameter to change the sign of the effective mass of
the ''relativistic'' electrons. We show that this TQPT is a third order
quantum phase transition, while most other conventional continuous quantum
phase transitions are second order. We also analyze the effect of finite
temperature on this TQPT, we find that there is no phase transition at $T>0$%
, in another word, there is no finite temperature phase transition
for the Anomalous Hall Effect(AHE). The temperature drives the TQPT
to be a crossover. The critical behavior as well as the properties
of correlation function in the finite temperature have also been
analyzed, then an experiment is proposed to detect these
unconventional properties. When there exists short range
interaction, we deal with this problem by the method of mean field
and show that this TQPT is robust against the weak short range
interaction for either the repulsive or attractive case. This is
very different from the regular Fermi liquid. Strong interaction
would change the pictures: strong enough repulsive interaction would
lead to a charge density wave (CDW) state\cite{Herbut}, while strong
attractive interaction would result in a superconductor
\cite{Uchoa}. To make our above results applicable to the orginal
Haldane model, we investigate the effect of the on-site energy
,which shifts the degeneracy of the two fermi points. Furthermore,
we point out the difference between this kind of TQPT and the
disorder-induced localization-delocalization transition in the
conventional quantum Hall systems\cite{Huckestein}, as well as in
the quantum spin Hall effect\cite {Onoda, Senthil}. Finally, we show
that though the Haldane model could not be realized in the present
experiment, our results can be applied to a large class of TQPT
which could be realized in experiment.

First, we briefly review the Haldane model\cite{Haldane,Yakovenko} and focus
on the phase transition in it. Haldane considered a (2+1)-dimensional
((2+1)-D) model in which spinless fermions hop on a half-filling honeycomb
lattice and couple with a periodic external magnetic field. There is no
''net'' magnetic flux through a unit cell and thus there is no Landau
quantization. However, a next nearest-neighbor hopping term with a phase $%
t_2e^{i\Phi }$ is introduced in the honeycomb lattice, which is a
time-reversal breaking term and leads to a ''chiral'' fermions without the
fermion doubling effect. This system provides a condensed-matter analog of
(2+1)-dimensional electrodynamics\cite{Semenoff} due to its linear energy
spectrum around the Fermi point.

In the honeycomb lattice, noninteracting fermions behave as a semimetal with
the Fermi surface shrunk into to two inequivalent points in one unit cell as
shown in Fig. 1. The general action of (2+1)-D noninteracting fermions in a
uniform or periodic potential is given by
\begin{equation}
S=Tr\int \frac{d^3k}{{(2\pi )}^3}\Psi ^{\dag }(k)G^{-1}(k)\Psi (k).
\end{equation}
For the Haldane model, it is convenient to represent $G^{-1}$ in a basis of
spinors $\Psi _k=(\Psi _{kA},\Psi _{kB})^T$ with A and B representing the
different sublattice of the honeycomb lattice. Therefore we have
\begin{figure}[tbh]
\includegraphics[width=3.4in] {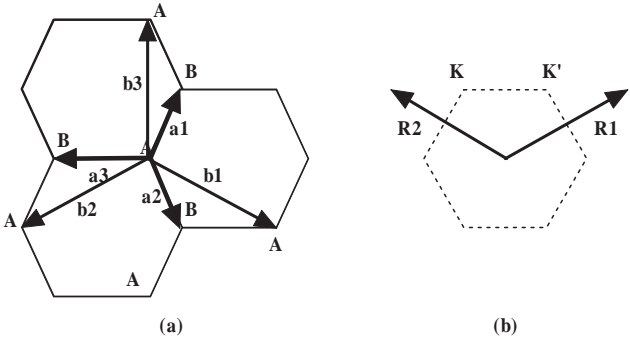}
\caption{(a)The honeycomb lattice as a superposition of two sublattice A,B.
The basis vectors are $\mathbf{a_1,a_2,a_3}$, two sublattice are connected
by $\mathbf{b_1,b_2,b_3}$. (b) The Brillouin zone, $\mathbf{R_1,R_2}$ are
basis vector in reciprocal-lattice. $\mathbf{K,K^{\prime }}$ are two
distinct fermi point in one Brillouin zone}
\label{fig1}
\end{figure}
\begin{equation}
G^{-1}=\mathbf{I}[i\omega +2t_2\cos \Phi \sum_i\cos (\mathbf{k\cdot b_i})]+%
\mathbf{d\cdot \sigma },
\end{equation}
where $\mathbf{d\cdot \sigma =}\sum_id_i\mathbf{\sigma }^i$ with $%
d_1=t_1\sum_i\cos (\mathbf{k\cdot a_i})$, $d_2=t_1\sum_i\sin
(\mathbf{k\cdot a_i})$ , $d_3=-t_2\sin \Phi \sum_i\sin
(\mathbf{k\cdot b_i})$, and the Pauli matrices $\sigma ^i(i=1,2,3)$
. Here we set the on-site energy $M$ in the original Haldane model
to be zero for convenience, we would discuss the effect of $M$ in
Section V. $t_1$ and $t_2e^{i\Phi }$ are the tunnelling amplitudes
between the nearest and the next-nearest neighboring sites. The
parameters $\mathbf{a_1,a_2}$ and $\mathbf{a_3}$ are the
displacement from one site to its nearest neighborhoods and $\mathbf{%
b_1=a_2-a_3}$, $\mathbf{b_2=a_3-a_1}$, etc. We could calculate the Chern
Number (Hall conductance) for this model using the standard representation%
\cite{Qi}
\begin{equation}
\sigma _{xy}=\frac 1{8\pi ^2}\int_\Omega d^2k[\mathbf{n}\cdot \partial _x%
\mathbf{n}\times \partial _y\mathbf{n}]
\end{equation}
with $\mathbf{n}=\mathbf{d}/|\mathbf{d}|$. Finally, we could obtain
\begin{equation}
\sigma _{xy}=\left\{
\begin{array}{lll}
1 &  & \text{if}\quad t_2\sin \Phi >0 \\
-1 &  & \text{if}\quad t_2\sin \Phi <0 \\
0 &  & \quad \text{otherwise}
\end{array}
\right. .
\end{equation}
The Chern number is the topological quantum number\cite{Thouless}. It was
first introduced into the condensed matter physics in the integer quantum
Hall effect to explain the stability of the Hall conductance to weak
pertubations\cite{Niu}. The physical meaning of the Chern number is the
quantum Hall conductance in our case. Different Chern numbers characterize
different topological phase and we focus on the phase transition between
them.

This phase transition could be interpreted in a traditional way by
calculating the singularly of the ground state energy.. The energy spectrum
for (2) is :
\begin{equation}
E(m,\mathbf{k})=\pm \sqrt{d_1^2+d_2^2+d_3^2}
\end{equation}
where $d_i$ have been defined above, and $m=t_2\sin (\Phi )$. In the ground
state, the lower band is filled completely so the ground state energy
\begin{equation}
E(m)=\int \int_\Omega d^2\mathbf{k}\quad E(\mathbf{k},m)
\end{equation}

\begin{figure}[tbh]
\includegraphics[width=3.5in] {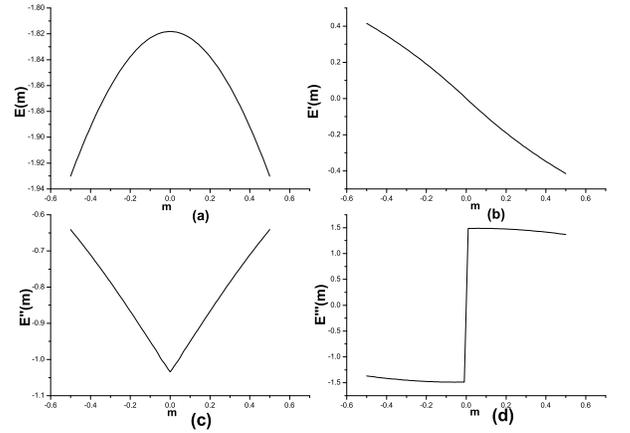}
\caption{The ground state energy(a) and its first order derivative(b),
second order derivative(c), third order derivative (d)}
\label{fig2}
\end{figure}

To find the singularity of $E(m)$ in the phase transition point $m$=0, we
calculate $E(m)$ and its derivative, the numerical result is shown in Fig.2.
We can see that the second deviative is continuous while the third one is
not in the phase transition point, which means that this quantum phase
transition is the third order quantum phase transition. This result could be
explained in a heuristic way when we consider the behavior around the two
distinct independent Fermi points. These two independent effective
Hamiltonians $H_\alpha $ ($\alpha =1,-1$ represent different Dirac Fermi
points) are given by

\begin{equation}
H_\alpha =\mathbf{D}_\alpha \cdot \mathbf{\sigma }
\end{equation}
where $\mathbf{D}_\alpha =(\alpha k_yc,k_xc,m_\alpha )$, $c=3/2t_1$, $%
m_\alpha =-3\sqrt{3}\alpha t_2\sin \Phi $. The parameter $c$ is the ''light
velocity'' in the relativistic Hamiltonian and below we set $c=1$ for
simplicity.

The total Hamiltonian is\cite{Haldane} :
\begin{equation}
H=\Psi _{-1}^{\dag }H_{-1}\Psi _{-1}+\Psi _1^{\dag }H_1\Psi _1.
\end{equation}
$\Psi _{\pm 1}$ represent fermions around different fermi points. When we
adjust the parameter $\Phi $ from negative to positive, the mass of both
Dirac fermions change sign and the phase transition occurs.

Because the two Dirac Fermi points are independent with each other, below we
only focus on one of them. From the Hamiltonian (7), we can easily calculate
the energy spectrum $E(\mathbf{k})=\pm \sqrt{\mathbf{k}^2+m^2}$ with $+$ for
the ''conduction'' band and $-$ for the ''valence'' band. The energy of the
ground state could be estimate as
\begin{equation}
E(m)=-\int \int d^2\mathbf{k}\sqrt{\mathbf{k}^2+m^2}=-|m|^3+C.
\end{equation}
$C$ is the physical cutoff of the integral and is obviously analytic in the
phase transition point $m=0$, because we only concern about the singularity
of $E(m)$ at the critical point $m=0$, $C$ is unimportant. From (9) we find
that the third order derivative of $E(m)$ to $m$ is discontinuous in the
point $m=0$, so this TQPT is a third order phase transition. The analysis
above implies that Hamiltonian (8) provides a effective approximation of the
actual Hamiltonian in eq. (1) if we only concern the behavior around the
Fermi point. Below we use eq.(8) to analyze the properties of the phase
transition for simplicity.

\section{Finite temperature effect}

A typical example of finite temperature effect on the quantum phase
transition is the transverse Ising Model\cite{Sachdev}, which is fermionized
by the Jordon-Wigner transformation. The finite temperature drive the
quantum phase transition into a crossover. It is shown that there is a
similar behavior in our case. When the temperature is not too high, the
model could be expressed in terms of a continuum canonical fermion field $%
\Psi =(\Psi _{\uparrow },\Psi _{\downarrow })^T$. Its partition function is
\begin{equation}
Z_F=\int D\Psi ^{\dag }(\mathbf{x},\tau )D\Psi (\mathbf{x},\tau
)exp(-\int_0^{1/T}d\tau \int d^2\mathbf{x}L_F)
\end{equation}
with
\begin{eqnarray}
L_F &=&\Psi _{\mathbf{-1}}^{\dag }(\frac \partial {\partial \tau }+\partial
_x\sigma ^y+\partial _y\sigma ^x-m\sigma ^z)\Psi _{\mathbf{-1}} \\
&&+\Psi _{\mathbf{1}}^{\dag }(\frac \partial {\partial \tau }+\partial
_x\sigma ^y-\partial _y\sigma ^x+m\sigma ^z)\Psi _{\mathbf{1}}.  \nonumber
\end{eqnarray}
Here $\tau $ is the imaginary time. From (10) and (11), we perform the
standard scaling transformation by rescaling lengths, times, as well as the
field etc . We define $x^{\prime }=xe^{-l},\tau ^{\prime }=\tau e^{-zl},\Psi
^{\prime }=\Psi e^l$ with $e^{-l}$ being the dimensional rescaling factor.
To make sure the new $L_F$ having the same form in terms of $x^{\prime
},\tau ^{\prime },\Psi ^{\prime }$, we find that $z=1$ and $m^{\prime }=me^l$%
. So we have $\dim [m]=1$ and the critical exponent $\nu =1$.

\begin{figure}[tbh]
\includegraphics[width=3.5in] {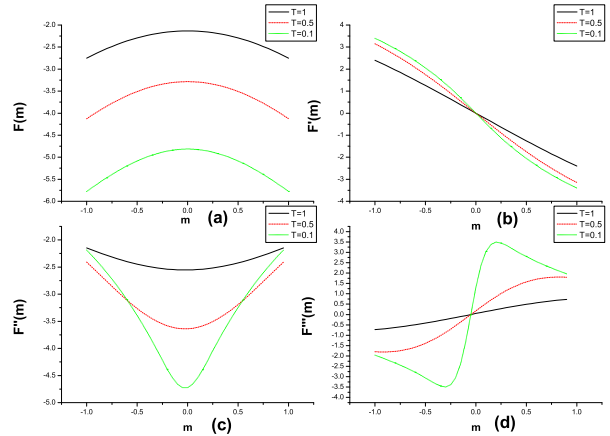}
\caption{The free energy(a) and its first order derivative(b), second order
derivative(c), third order derivative (d) at different temperature}
\label{fig3}
\end{figure}

At finite temperature $T$, the free energy $F$ is given by
\begin{equation}
F=-T\int \frac{d^2\mathbf{k}}{(2\pi )^2}2\ln (\cosh (\sqrt{\mathbf{k}^2+m^2}%
/T)).
\end{equation}
To find out whether there is a finite temperature phase transition, we
should calculate the derivative of the free energy. Numerical results are
shown in Fig.3. (Because we only consider the singularities around the phase
transition point, we set a physical cutoff of the integral and it would not
effect the singularities of (12).) We can see that the free energy is
analytic there is no singularity in (12) at $T\neq 0$, thus no phase
transition occurs. However, there is a crossover just like in the transverse
Ising model at finite temperature. The phase diagram is shown in Fig.2.
There are two regions in the phase diagram: the low-T massive Chiral-liquid
and the continuum high-T region (quantum critical region). However, the
temperature could not be too high that the continuum action $L_F$ in (11)
fails to provide a good approximation of the real system. Below we analyze
the properties in both regions by calculating the fermion Green's function: $%
G_F(\mathbf{x},t)=\langle \Psi (\mathbf{x},t)\overline{\Psi }(0,0)\rangle .$
In the quantum field theory, the Green's Function for the massive Dirac
Fermions at finite temperature is :
\begin{equation}
G_F^{ab}(\mathbf{x},t)=(i\gamma ^\mu \partial _\mu +m)_{ab}g_F(\mathbf{x},t)
\end{equation}
\begin{equation}
g_F(\mathbf{x},t)=-\int \frac{d^2\mathbf{k}}{(2\pi )^2}\frac 1{\sqrt{\mathbf{%
k}^2+m^2}}\frac{e^{-i\mathbf{k\cdot x}+i\sqrt{\mathbf{k}^2+m^2}t}}{1+e^{%
\sqrt{\mathbf{k}^2+m^2}/T}}.
\end{equation}
We use the 2D representation of $\gamma $ matrix ($\mu $=0,1,2) rather than
the ordinary 4D representation. $a,b$ is the spin of the Dirac electron. $t$
denotes the real time.

In our case, there are two kinds of Dirac fermions($\Psi_1$ and $\Psi_{-1}$)
and both of them contribute to the Green function, the Green function is
sightly different from the standard formula (13), as we will show below.
\begin{figure}[tbh]
\includegraphics[width=3.4in] {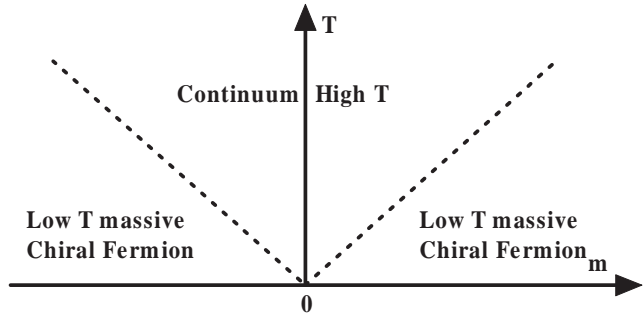}
\caption{Finite -T phase diagram of the TQPT as a function of the mass $%
\mathbf{m}$ of the Dirac fermion and the temperature T}
\label{fig4}
\end{figure}

\textit{Low-T massive Dirac-liquid, $T\ll |m|$}: In this region, the
temperature is very low and thus the physics is controlled primarily by the
critical line $T=0$. First we analyze $T=0$ equal-time correlations given by

\begin{equation}
g_F(\mathbf{x},0)=-\int \frac{d^2\mathbf{k}}{(2\pi )^2}\frac{e^{-i\mathbf{%
k\cdot x}}}{\sqrt{\mathbf{k}^2+m^2}}=\int_0^\infty \frac{dk}{2\pi }\frac{%
kJ_0(k|x|)}{\sqrt{k^2+m^2}}.
\end{equation}
We could approximately estimate the asymptotic behavior when $|\mathbf{x}%
|\rightarrow \infty $ by the contour integration which picks up contribution
from the poles at which function is divergent in the complex $k$ plane. For
large $x$, those poles closest to the real axis provide the dominant
contribution to the integral. The leading result is $G(|\mathbf{x}%
|,0)\approx $ $C(x)e^{-|m||\mathbf{x}|}$ ($|\mathbf{x}|\rightarrow \infty $%
), and thus the equal-time correlation exponentially decays in the spacial
dimension with a correlation length $\xi =1/|m|$.

Then we calculate the equal-space correlation. For $T=0$, we have
\begin{equation}
g(0,t)=-\int \frac{d^2\mathbf{k}}{(2\pi )^2}\frac{e^{i\sqrt{\mathbf{k}^2+m^2}%
t}}{\sqrt{\mathbf{k}^2+m^2}}.
\end{equation}
Doing this integration directly, we get
\begin{equation}
g(0,t)\varpropto \frac 1{it}e^{i|m|t}.
\end{equation}
So we find :
\begin{equation}
G^{AA}(0,t)\varpropto -\frac 1{it^2}e^{i|m|t}+\frac{|m|}te^{i|m|t}.
\end{equation}
We notice that the term $m$ in (13) doesn't appear in (18), because the $m$
in different fermi points have opposite sign and cancel each other so $%
G(0,t) $ is independent of the sign of $m$, which is reasonable in physics.

So $G(0,t)$ decays with a power law and oscillatory with a frequency $|m|$.
The correlation length in the real time axis (coherent time) is infinite
when $T=0$. We can see that though there is a gap, quantum systems at $T=0$
indeed have a long-range phase correlation in time which could not be seen
when we map the D-dimensional quantum system to the D+1 dimensional classic
system \cite{Sachdev}. When the temperature is very low, the density of the
quasiparticles is small, so the life of a quasiparticle (coherent time)
holds for a long time in this region.

Now we discuss the observable effect of $G(0,t)$ in this region. We
use the tunnelling effect in a bilayer Haldane model. There is a
voltage $V$ between the bilayer. By calculating the I-V curve in the
our system, we can find that it is totally different from the
ordinary electron systems\cite{Wen}.

Notice that we only consider the long-term behavior of $G(0,t)$ in
experiment, that is $t\gg \frac 1{|m|}$ (actually it is $t\gg \frac \hbar
{|m|}$, we set $\hbar =1$ in our case). So $|m|/t\gg 1/t^2$, the term $\frac
1{it^2}e^{i|m|t}$ is neglectable in (18), so $G^{AA}(0,t)\varpropto \frac{|m|%
}te^{i|m|t}$ . The correlation of the tunneling operator $I(t)=C_{1A}^{\dag
}(0,t)C_{2A}(0,t)$ is :
\[
\langle I(t)I^{\dag }(0)\rangle =-G_1^{AA}(t,0)G_2^{AA}(-t,0)\varpropto
\frac{m^2}{t^2},
\]
where 1, 2 denote the different layer and A, B are different sublattices in
each layer. Using the linear response formation, it is easy to find that the
tunneling current $I\varpropto m^2V$, which is different from the massless
case $I\varpropto V^3$.

\begin{figure}[tbh]
\includegraphics[width=3.5in] {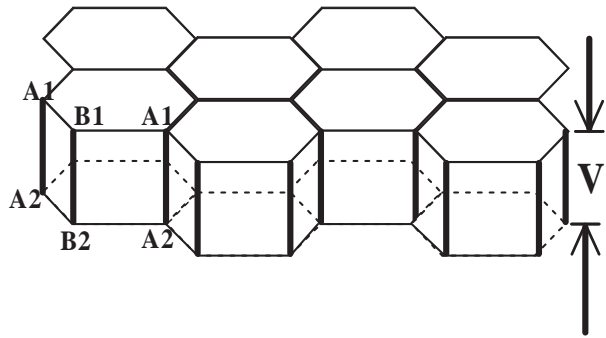}
\caption{ The bilayer Haldane model with voltage V between the layers. The
I-V curve of the tunneling current could be measured in this device}
\label{fig5}
\end{figure}

\textit{Continuum high T region, $T\gg |m|$}: This region is also known as
the ''quantum-critical region'', which plays an important role in condensed
matter physics because its potential relation with the heavy fermion systems%
\cite{Steglich} and the high-Tc superconductivity\cite{Varma}. The de
Broglie wavelength of the excitation in this region is the same order as
their mean spacing, so quantum and thermal effect entangle and both of them
play an equally important role. We mainly focus on the line $m=0$, and
calculate the equal-space correlation to see how the effect of temperature
change the behavior (13). From

\begin{equation}
g(0,t)=-\int \frac{d^2\mathbf{k}}{(2\pi )^2}\frac{e^{i|k|t}}{%
|k|(1+e^{-|k|/T})}=\int_0^\infty \frac{dk}{2\pi ^2}\frac{e^{ikt}}{1+e^{-k/T}}%
,
\end{equation}
This integral could be done by the contour integration. In the experimental
condition: $t\gg \hbar /k_BT$ (we set $k_B=\hbar =1$ in our case), it is
easy to find that
\begin{equation}
G(0,t)\varpropto T^2e^{-\pi tT}.
\end{equation}
We can see that (21) is totally different from (18). The equal-space
correlation decays exponentially and the phase coherent time is proportional
to $1/T$ in this region.

\section{Short range interaction}

For fermions, the interaction and interaction-induced quantum phase
transition play a central role in condensed matter physics\cite
{Hertz,Shankar}. We will show that our TQPT is stable against both the weak
repulsive and attractive interaction, while strong enough
repulsive/attractive interaction would lead to the CDW/superconductor state.

First we consider the condition in the critical point of the TQPT $(m=0)$.
The Hamiltonian for the spinless fermions in the honeycomb lattice with
nearest-neighbor interaction $V$ is given by
\begin{equation}
H=\sum_i\sum_{\mathbf{a}}[-\frac 12(a_i^{\dag }b_{i+\mathbf{a}%
}+h.c)+V(n_i^a-\frac 12)(n_{i+\mathbf{a}}^b-\frac 12)],
\end{equation}
where $\mathbf{a}$ is the displacement from one site to its nearest neighbor
site. For the repulsive interaction, we use the standard mean-field to deal
with the interaction. Assuming $\langle n_i\rangle =1/2+1/2(-1)^i\Delta $
with $i=0\quad $for $i\in A$ and $i=1$ for $i\in B$, we get
\begin{equation}
H=\sum_i\sum_{\mathbf{a}}[-\frac 12(a_i^{\dag }b_{i+\mathbf{a}}+h.c)+V(\frac{%
\Delta ^2}4-\Delta (-1)^in_i)].
\end{equation}
Following the standard procedure\cite{Shankar}, we calculate the energy of
the ground state. Minimize it with respect to $\Delta $, we obtain the
self-consistent equation
\begin{equation}
1=V\int \frac{d^2\mathbf{k}}{(2\pi )^2}\frac 1{\sqrt{E^2(\mathbf{k})+\Delta
^2V^2}}
\end{equation}
with $E^2(\mathbf{k})=4\cos ^2(k_x/2)+4\cos (\sqrt{3}k_y/2k_y)\cos (k_x/2)+1$
at the critical point ($m=0$). When we deviate from the critical point,
there is a mass term in $E^2(\mathbf{k})$, and it is estimated that $E^2(%
\mathbf{k},m)\approx 4\cos ^2(k_x/2)+4\cos (\sqrt{3}k_y/2)\cos
(k_x/2)+1+m^2. $

By solving this equation, we find that there is a SM-CDW transition at a
finite value $V_c(m)=0.98$. The numerical result is shown in Fig.6. For the
weak interaction $V<V_c$, $\Delta =0$, and the TQPT is stable. It is
consistent with previous study of the chiral symmetry breaking induced by
Coulomb interaction in Layered Graphite\cite{Khveshchenko}. For attractive
interaction, a mean field similar to the BCS theory could be done. The
SM-Superconductor phase transition in the honeycomb lattice has been analyzed%
\cite{Zhao,Uchoa}. These works support the result that, for the half-filling
fermions, there exist a SM-Superconductor phase transition in $V=-V_c$ and
when $V>-V_c$, the SM is stable. This result is also correct when $m$
deviates from 0 slightly, so our TQPT is stable for the weak attractive
interaction.

In addition, it is interesting to study the effect of the temperature in the
global phase diagram in Fig.6: In the CDW phase, it is known that there is a
finite-temperature phase transition for the 2D CDW system. When the
temperature is high enough, the CDW order is destroyed. However, for the
topological phase, as analyzed above, there is no thermodynamic phase
transition at finite temperature, this result has shown a critical
difference between the topological phase and non-topological phase.

\begin{figure}[tbh]
\includegraphics[width=3.4in] {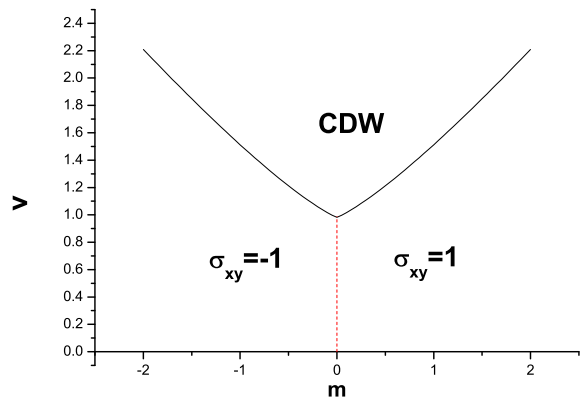}
\caption{The phase diagram of the TQPT with the next-nearest interaction $%
\mathbf{V}$, the solid line is the phase boundary between the CDW and
topolqiogical phases, and the dot line is the phase boundary of our TQPT }
\label{fig6}
\end{figure}

\section{Effect of the on-site energy M}
Up to now, We assume the on-site energy in the original Haldane
model $M=0$, as a consequence, the two kind of Dirac Fermions
$\psi_1$ and $\Psi_{-1}$ degenerate, the effective mass of them:
$m_\alpha =-3\sqrt{3}\alpha t_2\sin \Phi$. At the Phase transition
point, the mass of both kinds of fermions change sign simultaneously
thus the Chern number of the whole systme changes from $-1$ to $+1$.
When the on-site energy M is included, as in the original Haldane
model, the condition is different. Now the effective mass of the two
kind of Dirac fermions $m_\alpha =M-3\sqrt{3}\alpha t_2\sin \Phi$,
thus the energy of them no longer degenerate. The phase diagram with
$M$ has been demonstrated by Haldane (Fig.7). We assume $M>0$, as
shown in Fig.7, the original critical point $\Phi=0$  is split into
two different phase transition point $\Phi_{\alpha}=\alpha
\sin^{-1}(M/{3\sqrt{3}}t_2)$, $\alpha=\pm1$. At the point of
$\Phi_+$, for instance, only the mass of one kind of $\Psi_{-1}$
change sign, while the other is not, thus the Chern number of the
whole system is changed from $+1$ to $0$, different from the
condition when we set $M=0$. At this phase transition point, we will
find that only one kind of Dirac fermions experience the same
singularity as shown in Fig.2, while the energy of the other one is
analytic in this transition point. However, if we only focus on the
properties of the phase transition point, the analytic part is not
important, so all the results in above sections can be applied to
this kind of phase transition,  that the it is a third order quantum
phase transition, the temperature would drive it to be a crossover
and the weak interaction can't change the TQPT.

\begin{figure}[tbh]
\includegraphics[width=3.5in] {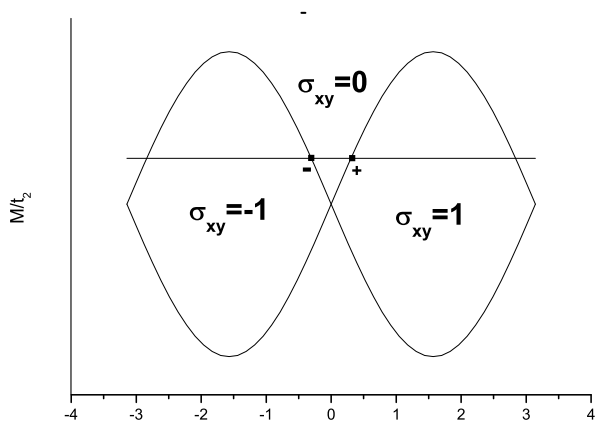}
\caption{The phase diagram in the original Haldane model. The
on-site energy $M$ shifts the degeneracy between the two fermi
points and splits our original TQPT into two TQPTs (+,$-$)}
\label{fig7}
\end{figure}

\section{Application and Discussion}

Now we discuss the important difference between our TQPT and one of the most
typical TQPT, i.e., the disorder-induced localization-delocalization
transition between different Hall plateaus in the Quantum Hall effect\cite
{Huckestein,Sondhi}. This kind of phase transition is controlled by the
impurity and it is, in nature, a kind of percolation phase transition with a
critical exponent $\nu =7/3$. In our TQPT, however, there are no impurities
and the phase transition is controlled by the mass of Dirac fermions. From
the analysis above, we can get that $\nu =1$. So our TQPT is totally
different from the phase transition in\cite{Huckestein}. It is also
different from the localization-delocalization transitions in the quantum
spin Hall effect\cite{Onoda} with $\nu =1.66$.

Since the Haldane model is an artificial model which could not be realized
easily in experiments, we would turn to other physical systems in order to
find this TQPT experimentally. In a recent work about quantum spin Hall
effect\cite{Bernevig,Dai}, an experiment in HgTe Quantum well is proposed to
realize this TQPT. This simplified model is based on the $\mathbf{k}\cdot p$
perturbation theory. For each kind of electrons with spin $\uparrow $ or $%
\downarrow $, there is a Dirac-type subband due to the special structure of
the quantum well. When the thickness of quantum well is adjusted to a
certain point, the effective mass of two kinds of Dirac electrons would
change its sign and a TQPT occurs. The system does not break the
time-reversal invariance, thus there is no Hall conductance, instead, there
exists a quantum jump of the spin Hall conductance $\Delta \sigma _{xy}^s=%
\frac{2e^2}h$. However, there is a difference between this TQPT and our
case. The mass term in \cite{Bernevig} is $\mathbf{k}$ dependent: $%
m=m^{*}-ck^2$, which makes their Hamiltonian different from the
Standard Dirac fermion. This difference results that the phase
transition is no longer between two symmetric phase with the
topological number $\sigma =\pm 1 $, but between a spin hall
insulator ($\sigma ^s=0$) and a spin hall conductance ($\sigma
^s=2$). However, if we only focus on the phase transition, simple
scaling analysis implies that the term $k^2$ is relevant in the
phase transition point, so our results in this paper could  be
applied to this TQPT.

It is natural to ask what happens if the interaction is long range (Coulomb
interaction), which corresponds to the more physical situation. Since the
weak short-range interaction is irrelevant in the critical point ($m=0$),
simple scale analysis would imply that the TQPT is still stable for the weak
long range interaction. However, when the interaction is strong enough, the
situation is complex and whether there is a phase transition driven by the
interaction is still controversial\cite{Son}. To answer this question, it is
necessary to do the renormalization group analysis of the (2+1)-Dimensional
massive QED coupling with the gauge field.

\end{document}